# A new temperature scale T* in lead-based relaxor systems

B. Dkhil[1*], P. Gemeiner[1], A. Al-Barakaty[2], L. Bellaiche[3], E. Dul'kin[4], E. Mojaev[4], M. Roth[4]

[1]*Laboratoire Structures, Propriétés et Modélisation des Solides, CNRS-UMR 8580, Ecole Centrale Paris, 92295 Châtenay-Malabry cedex, France*

[2]*Physics Department, Teacher College, Umm Al-Qura University, Makkah, Saudi Arabia*

[3]*Department of Physics, University of Arkansas, USA*

[4]*Faculty of Science, The Hebrew University of Jerusalem, Jerusalem 91904, Israel*

Via a combination of various experimental and theoretical techniques, a peculiar, identical temperature scale T* is found to exist in all complex lead-based relaxor ferroelectrics. T* corresponds to a nanoscale phase transition due to random fields. Interestingly, T* also exists in other oxides with extraordinary properties, such as giant magnetoresistivity or superconductivity. By analogy with such latter systems, the giant piezoelectricity related to relaxors might originate from proximity competing states effect.

*to whom correspondence should be addressed : brahim.dkhil@ecp.fr

Lead-based perovskite relaxors Pb(BB')O$_3$ such as Pb(Mg$_{1/3}$Nb$_{2/3}$)O$_3$ (PMN), Pb(Zn$_{1/3}$Nb$_{2/3}$)O$_3$ (PZN) or Pb(Sc$_{1/2}$Nb$_{1/2}$)O$_3$ (PSN) and related materials have recently attracted a great deal of attention, due to the excellent piezoelectric properties of their solid solutions with PbTiO$_3$ (PT) within the MPB[1]. In contrast to classical ferroelectric, relaxors display[2] *(i)* a strong frequency dependence of the dielectric permittivity, *(ii)* a broad dielectric anomaly around the temperature of the maximum dielectric permittivity, T$_m$, for a given frequency and *(iii)* no structural macroscopic phase transition across T$_m$. Nowadays it is generally accepted that the strong dielectric relaxation of T$_m$ needs two crucial ingredients. The first one concerns the quenched random electric and strain fields[3] (RFs) arising from chemical disorder and the difference in ionic charges and radii between the different kinds of B cations (in fact, the cations on the B-site of the perovskite structure typically deviate from a perfectly random distribution by possessing a short range order at a nanoscale through chemically ordered regions). The second ingredient concerns the dynamical polar nanoregions (PNRs), which nucleate several hundred degrees above T$_m$ at the Burns temperature[4], T$_B$. On cooling, the dynamic of the PNRs slows down and freezes out at the freezing temperature[5], T$_f$. In case of PMN, T$_B$, T$_m$ at 1kHz and T$_f$ are close to 630K, 260K and 220K respectively[4,5]. Another interpretation of T$_f$ is related to a phase transition into a nanodomain state due to the existence[3] of the RFs.

However, it was also suggested that another intermediate temperature exits. This temperature was first proposed to occur around 400K in the model relaxor PMN by Viehland *et al.* via dielectric constant measurement[6] and was considered to be a local Curie temperature. Later by means of diffraction data[7], Dkhil *et al.* suggested the existence of such intermediate temperature, and proposed that it corresponds to a local Heisenberg-Ising-like phase transition associated with an order-disorder transition of the $Pb^{2+}$ cations. Svitelsky *et al.* also proposed the existence of this new temperature (which they called T*) via an analysis of Raman spectra[8]. Despite some additional studies[9-12], the existence of T* is still controversial and therefore requires definite proofs. In this Letter, we undoubtedly demonstrate that the temperature scale T* is a definite signature of complex lead-based $Pb(BB')O_3$ relaxors and corresponds to a local phase transition that gives rise to the appearance of static polar nanoclusters. We demonstrate that T* is neither dependent on the B(B') species nor on the amount of doping (with Ti), unlike other critical temperatures (such as the macroscopic Curie and Burns temperatures). Moreover, theoretical calculations show that T* exists if RFs through chemical disorder are present and, in agreement with our experimental observations, T* is not affected by the strength of these RFs unlike the ferroelectric transition. Finally, by analogy with other nanoscale inhomogeneous systems[13,14], we propose that the giant piezoelectricity observed in relaxor-based systems within their morphotropic phase boundary (MPB) can be explained by phase separation and proximity competing effects. This work brings a new microscopic view of relaxors-based materials and new insights for a universal picture of materials with colossal effects.

Figure 1 displays the acoustic emission (AE) radiation for PMN and PZN from 450K to 750K. Experimental details for this technique can be found in Ref.[11]. There are two acoustic signals (thus associated with a strain release) that can be seen at around 500K and 630K for PMN versus 500K and 740K for PZN. 630K and 740K correspond to the Burns temperature for PMN and PZN, respectively, as consistent with previous works, whereas we propose that 500K is this new T* temperature. Interestingly, T* is the same for PMN and PZN, while these two compounds are rather different: PZN displays a macroscopic phase transition at $T_C \sim 400K$ (close to $T_f$) and has a $T_B \sim 740K$ [Ref. 4] whereas in PMN no phase transition occurs down to 5K and $T_f \sim 220K$ and $T_B \sim 630K$ [Refs. 4, 5].

Let us now check if the $T_B$ and T* temperatures can also be evidenced by x-ray diffraction, a technique which presents the practical advantage to be applied to powder or single crystal in small amount whereas AE requires a large enough single crystal. Two distinct changes are detected on the lattice parameter above 400K (fig. 2a for PMN and 2b for PZN): the most obvious one concerns the appearance of a plateau below T* whereas $T_B$ is associated with a weak change of slope. Figure 2 also shows the temperature dependence of the lattice parameter for several other Pb(BB')O$_3$ relaxors. It is remarkable that whereas $T_B$ can differ from one compound to another, T* remains close to 500K. Figure 2c further shows data obtained on partially chemically-ordered and disordered PSN, and reveals that T* and $T_B$ are independent on the B-cation ordering (unlike other critical temperatures such as $T_f$ and $T_C$ [15]). Furthermore, Fig. 2d reports x-ray data on PMN mixed with PT (PMN-PT) for different amount of mixing. It is obvious that the addition of PT does neither affect $T_B$ nor T*, while such addition is known to result in both an increase of the Curie phase transition temperature and a change in the low temperature ground state[11,12]. Such independency of $T_B$ and T* to PT was also observed in PZN-PT and PSN-PT (not shown here). Moreover, we also studied several other lead-based relaxors with different B-cations (size, charge, proportion). All of them show both a deviation around $T_B$ and T* (Figs 2e and 2f) and once again T* is the same among all the investigated compounds.

Let us now examine the Raman signal of several relaxors. For convenience, Figure 3 shows only a selected wavenumber region between 180 and 400 cm$^{-1}$ of the Raman spectrum of PMN. When one compares the spectra at different temperatures, it is easy to see that the mode at around 270cm$^{-1}$ becomes weaker and weaker when the temperature increases. We plotted on fig 3d the ratio of the intensity of the mode at 270cm$^{-1}$ (mode D) with respect to that at 240cm$^{-1}$ (mode C). Interestingly, mode D appears around T* indicating that a new "local" change takes place at this temperature. We also plotted on the same figure the same ratio but for PSN compound, which confirms that T* is the same for PSN and PMN (as consistent with our previous data).

Having undoubtedly demonstrated here that this new temperature T* indeed exists, let us look for a fingerprint of T* in previous works. For instance, such temperature corresponds to the maximum of the transverse acoustic phonon linewidth in Ref. [16], the crossover between low and high temperature of the transverse optical phonon wavelength in Ref. [17] and the appearance of the "butterfly-like" diffuse scattering in Ref. [18]. In other words, previous works further confirm the existence of T* even if they overlooked it!

Now let us turn our attention to the physical reason for the existence of T*. For that, it is important to emphasize again that our results clearly reveal that T* is the same for all lead-based systems studied here, unlike other critical temperatures. T* is thus not directly linked to the type of B-cations. Note that all the compounds we analysed show B-cations disorder that provides RFs. Therefore by affecting the already existing quenched RFs, no effect (if any) is detected on T* as the change of slope in the lattice parameter remains the same in any PMN-PT compound and in disordered or partially ordered PSN. Besides, it also appears that T* should be associated with lead $Pb^{2+}$ cations and/or oxygen network. $Pb^{2+}$ is known to be responsible for the main polarization in lead-based relaxors[7]. The fact that T* is a local Curie temperature[6] indicates that the nanoclusters becoming static at T* are polar. This polar character is also confirmed by recent calculations results [19,20]. As a matter of fact, calculations done by Iniguez and Bellaiche[19] showed that contrary to the normal ferroelectric PZT compound, the relaxor disordered PSN displays local polarization appearing far above the ferroelectric transition because of the existing RFs related to the B-cation disorder. Interestingly, they found that such polarization appears at $T/T_{max} = 1.2$ (where $T_{max}$ is the temperature at which the static dielectric response is maximum), which experimentally corresponds to T=480K in PSN and which is thus very close to the measured T*. Such theoretical work brings an additional support for the existence of T* and confirms that it is related to the existence of the quenched RFs. Interestingly when the RFs are strongly perturbed by removing a fair amount of Pb atoms[20], it results in an increased broadening of the static dielectric constant and a shift towards a lower temperature for the maximum of the static dielectric constant. However, the highest temperature at which the local polarization starts to develop nearly remained the same, that is T*. To confirm these results, we used the same calculations tool – that is, a first-principles-based effective Hamiltonian scheme -- to investigate disordered PSN (Sc and Nb are randomly distributed) and completely ordered PSN (planes of Sc atoms alternate with planes of Nb atoms along the [111] direction). Practically, periodic supercells with lateral size of 48 Å along any pseudo-cubic <001> direction are used in these simulations. Such nanometric cells can be thought as representing a single polar nanocluster with specific RFs in the case of disordered PSN, while no RFs exist inside the ordered PSN supercell (due to the B-cation arrangement). Figure 4 shows the local modes (that are directly proportional to the local polarization) as a function of $T/T_{max}$ for both disordered and completely ordered PSN supercells. As one can see, the local polarization is much broader in disordered, than in ordered, PSN. In particular, such local polarization starts to develop around $T/T_{max} = 1.2$ in the disordered PSN, while it is "only" significant for much lower ratio in ordered PSN. We interpret such difference as indicative of the fact that T* disappears when the RFs do not exist.

Our experimental and theoretical results thus show that T* is the temperature at which static polar nanoregions appear if RFs exist (and independently of the strength of the RFs above a certain, significant value). It is now of interest to confront our results for the relaxor systems to other nanoscale inhomogeneous systems with, e.g., CMR or HTSC and for which several evidences of existence of a peculiar temperature, T*, were reported[13,14,21,22]. It was suggested that T* is the result of the introduction of quenched disorder within several competing states and thus T* is a reminiscence of the clean limit transition between these ordering states. The temperature dependence was then schematically described in electronic systems as following. At high temperature, uncorrelated polarons appear at $T_{pol}$. These individual polarons start to correlate at T* becoming correlated nanoclusters. Thus at T* both uncorrelated and correlated polarons coexist. By decreasing temperature, the number and/or size of the correlated polarons increases and becomes maximum at $T_c$. This scheme can be exactly applied to relaxors by replacing $T_{pol}$ by $T_{Burns}$. It is also interesting to underline that in such strongly correlated electronic systems, colossal or proximity effect[13,14,21,23-25] was reported and that doping[14,21] does not affect T* whereas it affects $T_c$. Interestingly and as we have shown, in case of relaxor, T* is also not affected by the doping. Moreover relaxors also display a colossal effect that is the giant piezoelectricity.

By analogy with the electronic systems, it may be advocated that the giant piezoelectric response observed at the MPB of relaxors [26-30] is related to the competition between ordered states at a nanoscale in presence of disorder (that gives rise to RFs) and to the fact that any weak stress or electric field strongly affects such competition. It is well know that in ferroelectric perovskites one may consider several kind of competition between different symmetries (rhombohedral, tetragonal, monoclinic, …), different polar states (ferroelectric, antiferroelectric, paraelectric), different distortions (oxygen tilting, cation displacement) or different dynamics (order-disorder, displacive motions). Thus all the ingredients, i.e. competing phases and quenched disorder, are present in relaxors. Such result provides a new way to consider relaxors, that is as a member of a wide family of materials possessing nanoscale inhomogeneous states and exhibiting different colossal properties but all having common features as a function of temperature.

## References

1. B. Noheda, *Curr. Opin. Solid State Mater. Sci.* **6**, 27 (2002)

2. A.A. Bokov, Z.E. Ye, *J. Mater. Sci.* **41**, 31 (2006)

3. V. Westphal, W. Kleemann, M.D. Glinchuk, *Phys. Rev. Lett.* **68**, 847 (1992)

4. G. Burns, F.H. Dacol, *Phys. Rev. B* **28**, 2527 (1983)

5. D. Viehland, *et al.*, *Phys. Rev. B* **43**, 8316 (1991)

6. D. Viehland, *et al.*, *Phys. Rev. B* **46**, 8003 (1992)

7. B. Dkhil, *et al.*, *Phys. Rev. B* **65**, 024104 (2002)

8. O. Svitelskiy, *et al.*, *Phys. Rev. B* **68**, 104107 (2003)

9. B. Mihailova, *et al.*, *Phys. Rev. B* **77**, 174106 (2008).

10. O. Svitelskiy, *et al.*, *Phys. Rev. B* **72**, 172106 (2005)

11. E. Dul'kin, *et al.*, *Phys. Rev. B* **73**, 012102 (2006)

12. M. Roth, *et al.*, *Phys. Rev. Lett.* **98**, 265701 (2007)

13. E. Dagotto, *Science* **309**, 257 (2005)

14. E. Dagotto, *New Journal of Physics* **7**, 67 (2005)

15. F. Chu, I.M. Reaney, N. Setter, *J. Appl. Phys.* **77**, 1671 (1995)

16. A. Naberzhnov, *et al.*, *Eur. Phys. J. B* **11**, 13 (1999)

17. S. Wakimoto, *et al.*, *Phys. Rev. B* **65**, 172105 (2002)

18. G. Xu, *et al.*, *Phys. Rev. B* **69**, 064112 (2004)

19. J. Iniguez, L. Bellaiche, *Phys. Rev. B* **73**, 144109 (2006)

20. L. Bellaiche, *et al.*, *Phys. Rev. B* **75**, 014111 (2007)

21. J. Burgy, *et al.*, *Phys. Rev. Lett.* **87**, 277202 (2001)

22. M.B. Salamon, M. Jaime, *Rev. Mod. Phys.* **73**, 583 (2001)

23.     R.S. Decca, *et al.*, *Phys. Rev. Lett.* **85**, 3708 (2000)

24.     G. Alvarez, *et al.*, *Phys. Rev. B* **71**, 014514 (2005)

25.     C. Sen, G. Alvarez, E. Dagotto, *Phys. Rev. Lett.* **98**, 127202 (2007)

26.     S.E. Park, and T.R. Shrout, *J. Appl. Phys.* **82**, 1804 (1997)

27.     H. Fu, and R.E. Cohen, *Nature* **403**, 281 (2000)

28.     Z. Kutnjak, J. Petzelt, R. Blinc, *Nature* **441**, 956 (2006)

29.     R. Guo *et al.*, *Phys. Rev. Lett.* **84**, 5423 (2000)

30.     L. Bellaiche, A. Garcia, D. Vanderbilt, *Phys. Rev. Lett.* **84**, 5427 (2000)

**Figure legends**

**Fig.1** : Acoustic Emission (AE) activity from 450K to 750K for $PbMg_{1/3}Nb_{2/3}O_3$ (PMN) (black) and $PbZn_{1/3}Nb_{2/3}O_3$ (PZN) (red).

**Fig. 2** : Pseudo-cubic lattice parameter $a$ as a function of temperature from 300K to 850K obtained with the (200), (220) and (222) Bragg peaks for (a) $PbMg_{1/3}Nb_{2/3}O_3$ (PMN), (b) $PbZn_{1/3}Nb_{2/3}O_3$ (PZN), (c) $PbSc_{1/2}Nb_{1/2}O_3$ (PSN) partially ordered and disordered, (d) $PbMg_{1/3}Nb_{2/3}O_3-PbTiO_3$ (PMN-PT) with PT concentration of 0%, 10%, 25% and 33%, (e) $PbFe_{2/3}W_{1/3}O_3$ (PFW) and $PbFe_{1/2}Nb_{1/2}O_3$ (PFN), (f) $PbMg_{1/3}Ta_{2/3}O_3$ (PMT) and $PbSc_{1/2}Ta_{1/2}O_3$ (PST). The straight lines are guide to the eye.

**Fig. 3** : Selected wavelength Raman spectra between $180cm^{-1}$ and $400cm^{-1}$ at (a) 80K, (b) 270K and (c) 770K. Panel (d) represents the ratio between the intensities of mode C (around $240cm^{-1}$) and mode D (around $270cm^{-1}$).

**Fig. 4** : Local modes (i.e. quantity directly related to polarisation) as a function of $T/T_{max}$ for disordered PSN (full circles) and fully ordered PSN (open circles).

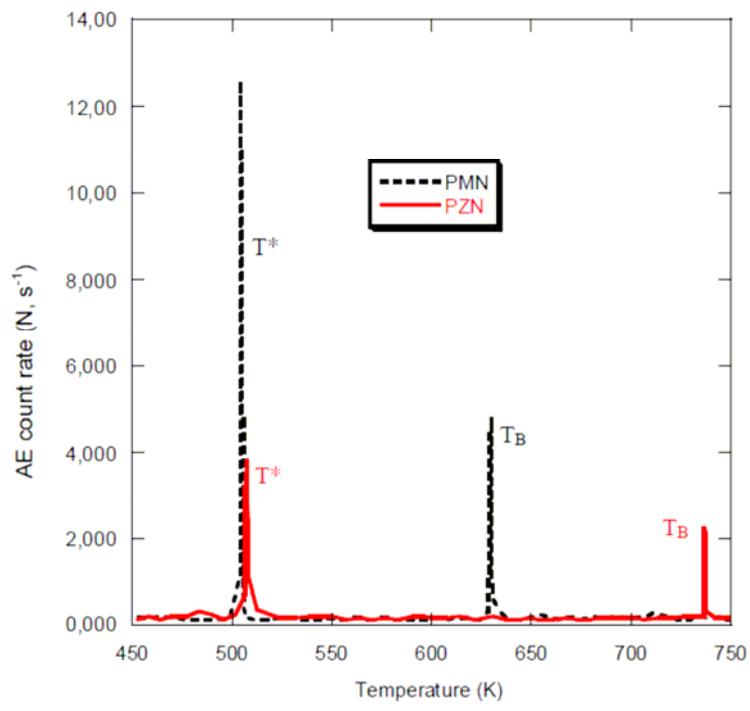

Fig. 1

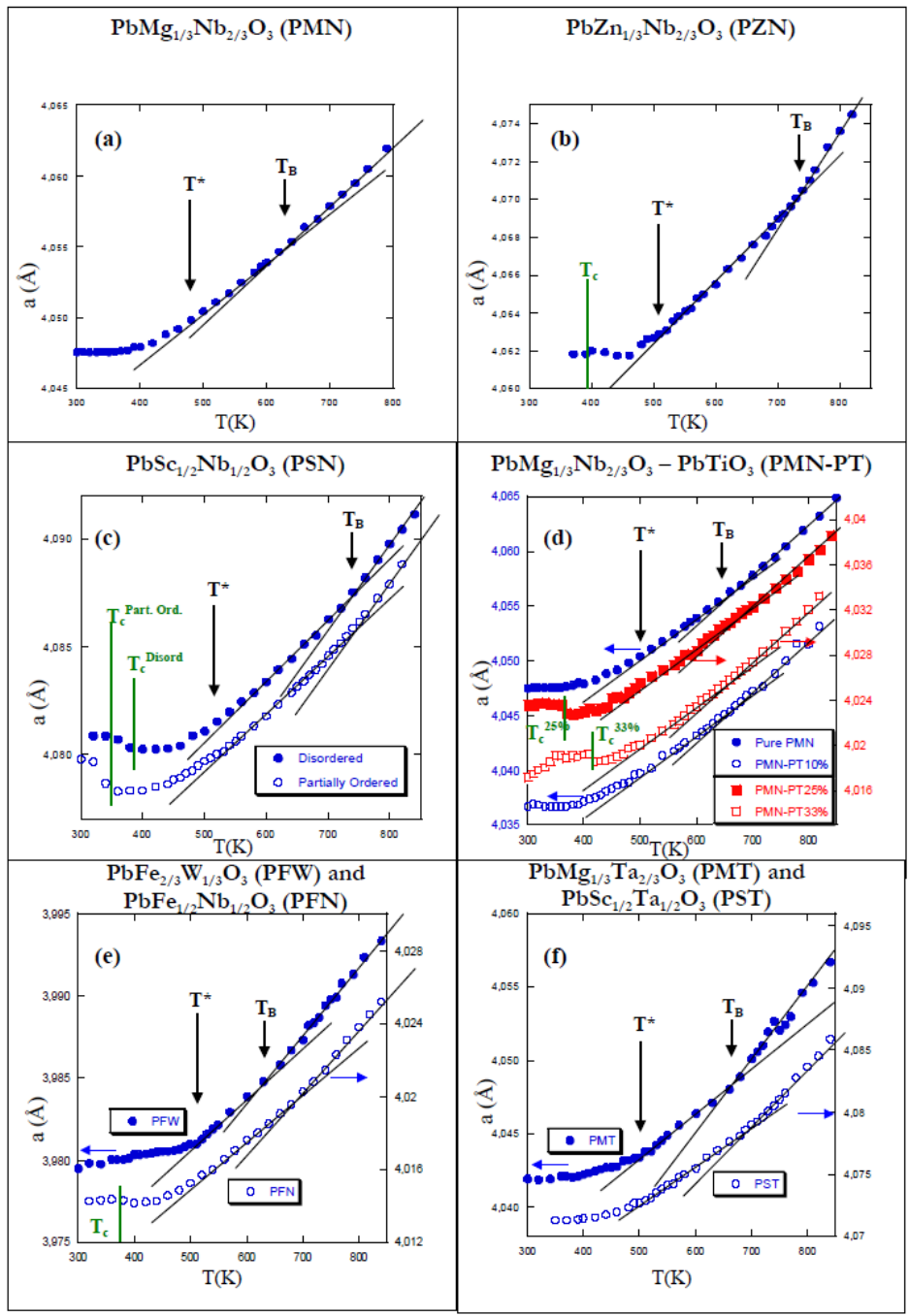

Fig. 2

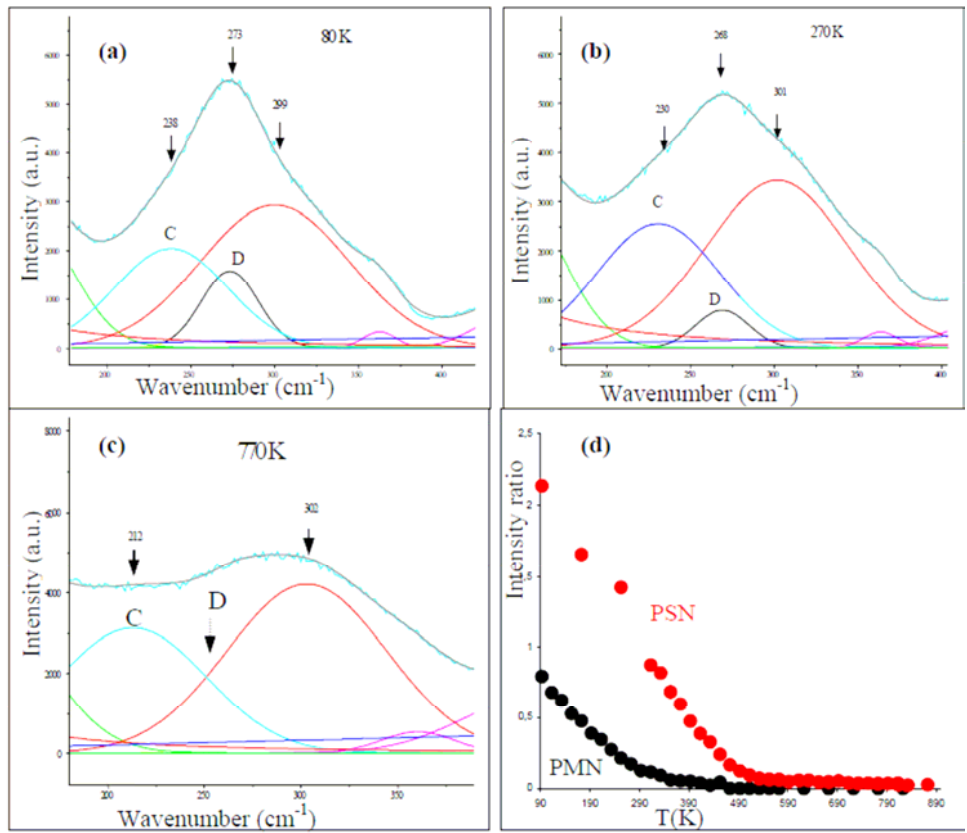

Fig. 3

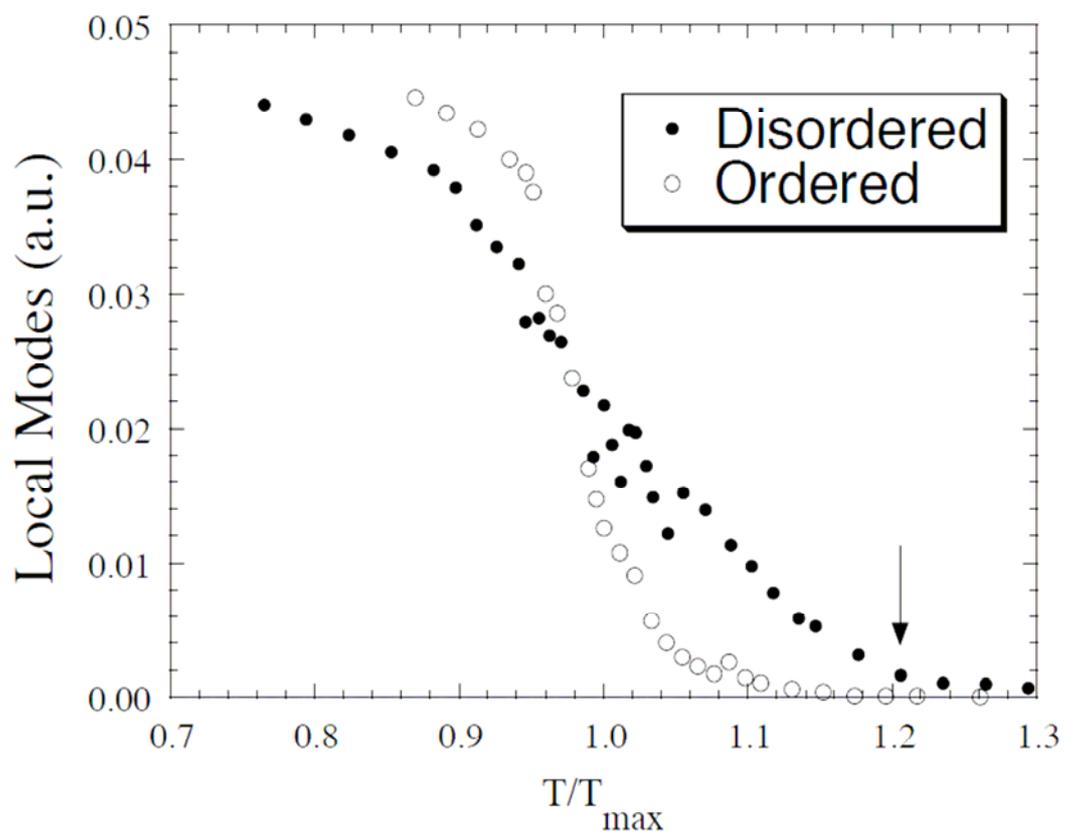

Fig. 4